\documentclass[floatfix,useAMS,usenatbib]{mnras}
\usepackage[greek,british]{babel}
\usepackage{amsmath}
\usepackage{graphicx,epsfig}
\usepackage{multirow}
\usepackage{verbatim}
\usepackage[figuresright]{rotating}
\usepackage{txfonts}
\usepackage{times}
\usepackage[T1]{fontenc}
\usepackage{aecompl}

\title[Modelling of Galactic GC VHE Fluxes and Their Uncertainties]{Assessing Uncertainties in the Predicted Very-High-Energy Flux of Globular Clusters in the Cherenkov Telescope Array Era}

\author[]{
Hambeleleni Ndiyavala-Davids$^{1,2}$\thanks{E-mail:hndiyavala@unam.na},
Christo Venter$^{1}$, Andreas Kopp$^{1}$,\newauthor
and Michael Backes $^{3,1}$\newauthor \\
\\
$^{1}$Centre for Space Research, North-West University, Potchefstroom Campus, Private Bag X6001, Potchefstroom 2520, South Africa \\
$^{2}$University of Namibia, Department of Science Foundation, Khomasdal Campus, Private Bag 13301, Windhoek, Namibia \\
$^{3}$University of Namibia, Department of Physics, Windhoek Campus, Private Bag 13301, Windhoek, Namibia}

\date{Accepted 2020 November 13. Received 2020 November 7; in original form 2020 October 19}
\pubyear{2020}
\begin{document}
\label{firstpage}
\pagerange{\pageref{firstpage}----\pageref{lastpage}}
\maketitle
\begin{abstract}

Terzan~5 is the only Galactic globular cluster that has plausibly been detected in the very-high-energy range. Stacking upper limits by H.E.S.S.\ on the integral $\gamma$-ray flux of a population of other globular clusters are very constraining for leptonic cluster emission models. We demonstrate that uncertainty in model parameters leads to a large spread in the predicted flux, and there are indeed regions in parameter space for which the stringent stacking upper limits are satisfied. We conduct two more case studies: we study the uncertainties in differential TeV flux for M15, showing that our model can satisfy the stringent MAGIC upper limits for this cluster, for typical cluster parameters. We also calculate the differential flux at TeV energies for $\omega$~Cen, from which five pulsars have recently been detected at radio energies. It is thus important to increase measurement accuracy on key model parameters in order to improve predictions of cluster fluxes so as to better guide the observational strategy of the Cherenkov Telescope Array.
\end{abstract}

%%%%%%%%%%%%%%%%%%%%%%%%%%%%%%%%%%%%%%%%%%%%%%%%%%
%%%%%%%%%%%%%%%%% BODY OF PAPER %%%%%%%%%%%%%%%%%%
\section{Introduction} \label{Introduction}
\citet{Ndiyavala2018} predict that tens of Galactic globular clusters (GCs) may be detectable within a reasonable amount of observation time for the next-generation Cherenkov Telescope Array (CTA). However, emission models predict quite a wide range of fluxes depending on how well constrained the model parameters are. 
%Models
There are a number of leptonic and hadronic models that predict the radiated spectrum by GCs. 
For example, \citet{Harding2005, Venter2005,Zajczyk2013} considered the cumulative pulsed curvature radiation (CR) spectrum from an ensemble of millisecond pulsars (MSPs) embedded in a GC due to leptons being accelerated within the pulsar magnetospheres. Furthermore,
\citet{Bednarek2007, Venter2008a,  Venter2009a, Venter2010, Cheng2010, Kopp2013,Zajczyk2013} all assume a leptonic origin of the unpulsed GC emission where pulsars inject accelerated leptons that radiate synchrotron radiation (SR) and inverse Compton (IC) emission as they traverse the GC. \citet{Ndiyavala2019} fit multi-wavelength data of Terzan~5 with a leptonic model that invokes unpulsed SR and IC components to model the radio and TeV data and cumulative pulsed CR to fit the \emph{Fermi} Large Area Telescope (LAT) data. The authors also explain the hard \emph{Chandra} X-ray spectrum via a ``new'' cumulative pulsed SR component from electron-positron pairs within the pulsar magnetospheres. Other models assume astrophysical objects such as white dwarfs that inject relativistic leptons into the GC \citep{Bednarek2012} or a $\gamma$-ray burst remnant that accelerates hadronic particles and secondary leptons, which contribute to the high-energy emission from GCs \citep{Domainko2011}. Recently, \citet{Brown2018} concluded that a combination of cumulative CR emission from MSPs and dark matter annihilation may explain the GeV emission detected by \emph{Fermi}-LAT from 47~Tucanae. 

The theoretical expectations above are motivated by broadband detections of GCs. The \emph{Fermi}-LAT has detected about $20$ $\gamma$-ray sources associated with GCs in the GeV band \citep{Abdo2009, Kong2010, Tam2011, Zhang2016}. In this paper, we focus on the very-high-energy (VHE; $E>100$~GeV) band. The ground-based High Energy Stereoscopic System (H.E.S.S.) Cherenkov telescope, operating in a pointing mode at energies above $100\,{\rm GeV}$, has only plausibly\footnote{The best-fit position of the source detected by H.E.S.S.\ is displaced from  Terzan~5's position, but still falls within the tidal radius of the GC. The morphology of the detected VHE emission is asymmetric and extends beyond the tidal radius of the GC, also being larger than the H.E.S.S.\ point spread function; however, the estimated probability of a chance coincidence of Terzan~5 and an unrelated VHE source is below $\sim10^{-4}$.} detected a single GC in our Galaxy, i.e., Terzan~5 \citep{Abramowski2011}. H.E.S.S.\ searched for VHE $\gamma$-ray emission from $15$ other Galactic GCs and did not detect any of those; neither did they detect any significant signal using a stacking analysis (involving a live-time-weighted GC flux; \citealt{Abramowski2013}.  Therefore, they derived integral flux upper limits from their single-GC and stacking analyses. Interpreting these upper limits, they noted that a simple scaling of existing theoretical predictions invoking IC from relativistic leptons produces a VHE flux that violates these limits\footnote{Indeed, the TeV flux is most sensitive to the particle source strength, i.e., the combination of number of MSPs, acceleration efficiency, and average MSP spin-down power, but not exclusively so. Thus, observational upper limits would first and foremost constrain the source strength, but may also be used to constrain degenerate parameters such as the diffusion coefficient or target photon fields (e.g., average stellar temperature), if the source strength is fixed to some reasonable value. If, however, limits are stringent enough that the resulting model parameters are not deemed reasonable anymore (e.g., too low number of MSPs or spin-down values given independent measurements from other wavebands), one would have to discard or extensively revise the model under consideration.}. Observations by other Cherenkov telescopes could only produce upper limits (e.g., \citealt{Anderhub2009, Mccutcheon2009}). Observations of the GC 47~Tucanae (NGC 104) were performed with H.E.S.S.\ leading to an upper limit on the integral $\gamma$-ray flux of $F(E>800\,{\rm GeV}) < 6.7 \times 10^{-13}\,{\rm cm^{-2} s^{-1}}$ \citep{Aharonian2009}. Observations of the GC M15 by the Major Atmospheric Gamma Imaging Cherenkov Telescopes (MAGIC) provided a deep upper limit on $F(E>300\,{\rm GeV})$, i.e., $<0.26\%$ of the Crab Nebula flux. Also, stringent differential flux upper limits were obtained on this source \citep{Acciari2019}. Although we focus on $\gamma$-rays, we note that GCs are broadband emitters, i.e., they are also detected in X-rays, optical, and  radio and these data are typically modeled using an SR spectrum \citep{Eger2010, Abdo2010, Clapson2011}.

The aim of this paper is to assess uncertainties in the predicted VHE $\gamma$-ray flux of GCs for a given leptonic model and to give theoretical guidance to CTA's observational strategy. That is, we model the IC $\gamma$-ray flux expected from several GCs to see whether the predicted flux exceeds the CTA sensitivity and whether we can satisfy the H.E.S.S.\ upper limits with a range of predicted fluxes (due to parameter uncertainties). We argue that one cannot simply scale the model output among different GCs, since the range of predicted fluxes depends crucially (and non-linearly in some cases) on the unique input parameters plus their uncertainties per source. The use of a population of clusters versus a single cluster should furthermore reduce the uncertainty in the predicted stacked flux, which can be used to make a better informed decision of whether a particular model should indeed be discarded in light of measurements or not; conversely, such an approach may yield more stringent constraints on the model parameters, if a given model remains viable. We will specifically (1) model a population of GCs and compare the predicted \emph{integral} fluxes with H.E.S.S.\ upper limits, (2) model both the  Northern-hemisphere  cluster, M15, and Southern-hemisphere one, Omega Cen ($\omega$~Cen), and compare our results with \emph{differential} flux upper limits. We chose these two GCs given the deep MAGIC upper limits found for M15, and given the recent radio pulsar detections in the case of the latter.
We will illustrate that there are indeed regions in parameter space for which the stringent stacking upper limits by H.E.S.S.\ are satisfied, for some subset of free parameters, given their uncertainty. Yet larger parameter uncertainties and more free model parameters will only increase the uncertainty on the predicted flux, making it even easier to satisfy these observational constraints.

The rest of the paper is structured as follows. In Section~2, we briefly discuss the model and its free parameters. In Section~3, we discuss the method employed to calculate the IC flux expected from the GCs plus its errors derived using a Monte Carlo process. Section~4 describes the results. In Section~5 we offer our conclusion.

\section{The Model and its Free Parameters} \label{sec:model} 
A large number of spectral components is expected from GCs \citep{Ndiyavala2019}. The unpulsed emission should transpire through the continuous injection of relativistic leptons by the MSPs into the ambient GC region \citep{Bednarek2007}, which in turn produce unpulsed SR and IC emission when they encounter the cluster magnetic field as well as background (stellar and cosmic microwave background) photon components. We also expect cumulative pulsed SR and CR components initiated by primaries and electron-positron pairs from within the magnetospheres of MSPs embedded in the cluster \citep{Harding2005, Venter2005,Zajczyk2013}.
In this paper, we focus on the VHE component. The model by \citet{Kopp2013} that we use here calculates the particle transport (including diffusion and radiation losses) in a spherically symmetric, stationary approach and predicts the spectral energy distribution (SED) from GCs for a very broad energy range by considering unpulsed SR and IC. The main free parameters of this model are \citep[cf.][]{Ndiyavala2018,Ndiyavala2019}: cluster magnetic field ($B$), power-law index ($\Gamma$) of the injected particle spectrum (the particles are assumed to have undergone acceleration in inter-pulsar shocks; \citealt{Bednarek2007}), number of GC stars ($N_{*}$),  distance to the cluster ($d$), the average spin-down luminosity per pulsar ($\langle\dot{E}\rangle$), the conversion efficiency of spin-down luminosity into particle acceleration ($\eta$), and number of MSPs in the GC ($N_{\rm MSP}$). The latter three are linked to the normalization of the injection spectrum $Q_{0}$ in the following way:
\begin{align}\label{eq:normalisation}
Q_{0} & =
\begin{cases}
\frac{L_{\rm eff}}{\ln(E_{\rm e, max}/E_{\rm e, min})}  & {\rm if} \quad \Gamma = 2.0,\\
\\
\frac{L_{\rm eff}(2-\Gamma)}{E_{\rm e, max}^{2-\Gamma} - E_{\rm e, min}^{2-\Gamma}} & {\rm if} \quad \Gamma \neq 2.0.
\end{cases}
\end{align}
where 
\begin{equation}\label{eq:eff_luminosity}
 L_{\rm eff} = \eta\langle\dot{E}\rangle N_{\rm MSP}
\end{equation}
is the effective luminosity, and $E_{\rm e,min}$ and $E_{\rm e,max}$ are the minimum and maximum particle energies, respectively. The model yields the differential VHE flux $dN_{\gamma}/dE$.
% \begin{equation}
%  \frac{dN_{\gamma}}{dE} = N_{0}E_{\gamma}^{-\alpha}.
%  \label{eq:diff_eq}
% \end{equation}
% In Equation~(\ref{eq:diff_eq}), $N_{0}$ is the flux normalisation and $\alpha$ is the spectral index.  
One can obtain the integral flux via
\begin{equation}\label{eq:integral_flux}
 F(E_{\gamma}\geq E_{\rm th}) = \int_{E_{\rm th}}^{E_{\gamma, \rm max}} E_\gamma\frac{dN_{\gamma}}{dE}dE_{\gamma}
\end{equation}
where $E_{\rm th}$ is the threshold energy for each cluster. We kept the spatial diffusion coefficient ($\kappa$), the core radius ($r_{\rm c})$, the half-mass radius ($r_{\rm h}$), and the tidal radius ($r_{\rm t}$) as fixed parameters. 
%For more details, see \citet{Ndiyavala2018, Ndiyavala2019}.

One should note that the free parameters of the model may be degenerate in the sense that changing several different parameters can independently produce higher TeV flux predictions. Moreover, while we treat them as independent parameters, some might actually be correlated, e.g., the cluster magnetic field may scale with the number of embedded pulsars in the cluster \citep[see,][]{Bednarek2007}. Since we do not know the exact nature of such a correlation, for simplicity we treat the mentioned free parameters as being independent. One may consider the effect of assuming some correlations between parameters, thereby decreasing the number of free parameters, in future.

\section{Monte Carlo Method}
\subsection{Integral Flux Upper Limits: the H.E.S.S.\ GC Population}
\label{sec:HESS_method}
\citet{Venter2015} applied a leptonic GC model to the population of clusters mentioned in \citet{Abramowski2013} to predict the live-time-averaged (stacked) flux and compare it to the H.E.S.S.\ stacking upper limits. They used fixed (best-guess) model parameters as noted in their Table~1, and used the inferred values from \citet{Abdo2010} for $N_{\rm MSP}$ where possible, and $N_{\star}$ values from \citet{Lang1993}, and obtained distances $d$ and structural parameters\footnote{http://gclusters.altervista.org/index.php} $r_{\rm c}$, $r_{\rm hm}$,
% used as a proxy for half-mass radius,
and $r_{\rm t}$ from \citet{Harris1996}. Typical values of $\eta\sim0.01$ \citep{Venter2005}, $N_{\rm MSP}\sim25$, and $\langle \dot{E} \rangle \sim 2 \times 10^{34}\,{\rm erg\,s^{-1}}$ led to typical values of $Q_{0}\sim10^{33-34}\,{\rm erg^{-1} s^{-1}}$ for the source strength (see Equations~[\ref{eq:normalisation}] and [\ref{eq:eff_luminosity}]). They also used $B = 5 \mu G$ and $\Gamma = 2.0$ throughout, and assumed optical photons from the stellar population embedded in the GC plus cosmic microwave background (CMB) photons as target fields for their IC calculation (assuming an average stellar temperature of $\langle T\rangle = 4\,500$~K). They found that none of the predicted single-cluster spectra violated the H.E.S.S.\ TeV upper limits for individual sources. Their averaged flux furthermore satisfied the stringent stacking upper limits. However, they noted that different choices of parameters led to average fluxes that in some cases may exceed these limits. That is why we refine their calculation in this paper, now also including an estimation of uncertainties on the predicted VHE fluxes.

In this Section, we describe the method we employ for modelling the IC $\gamma$-ray flux expected from the 15 Galactic GCs observed by H.E.S.S. We follow the method of \citet{Venter2008a} where they used a Monte Carlo technique to model the pulsed CR $\gamma$-ray flux and associated uncertainties expected from $100$ MSPs hosted by a GC. Their calculated CR flux was randomised over the magnetic inclination ($\alpha$) and observer ($\zeta$) angles that are usually estimated from radio polarisation measurements involving the rotating vector model \citep{Manchester1995}, as well as taking into account the fixed, pulsar-specific parameters such as period $P$ and period derivative $\dot{P}$ (or $\dot{E}$) of each MSP member. In our case, we use the same method to calculate the single-GC and stacked integral VHE flux of 15 GCs \citep{Abramowski2013}, with uncertainties due to the uncertainty in model parameters. We do this by randomising over seven free parameters as mentioned in Section~\ref{sec:model}: $B$, $\Gamma$, $N_{*}$, $d$, $\eta$, $\dot{E}$, and $N_{\rm MSP}$, in addition to fixing the structural and diffusion coefficient parameters as noted above.

\begin{figure*}
 \centering
 \includegraphics[width=.9\textwidth]{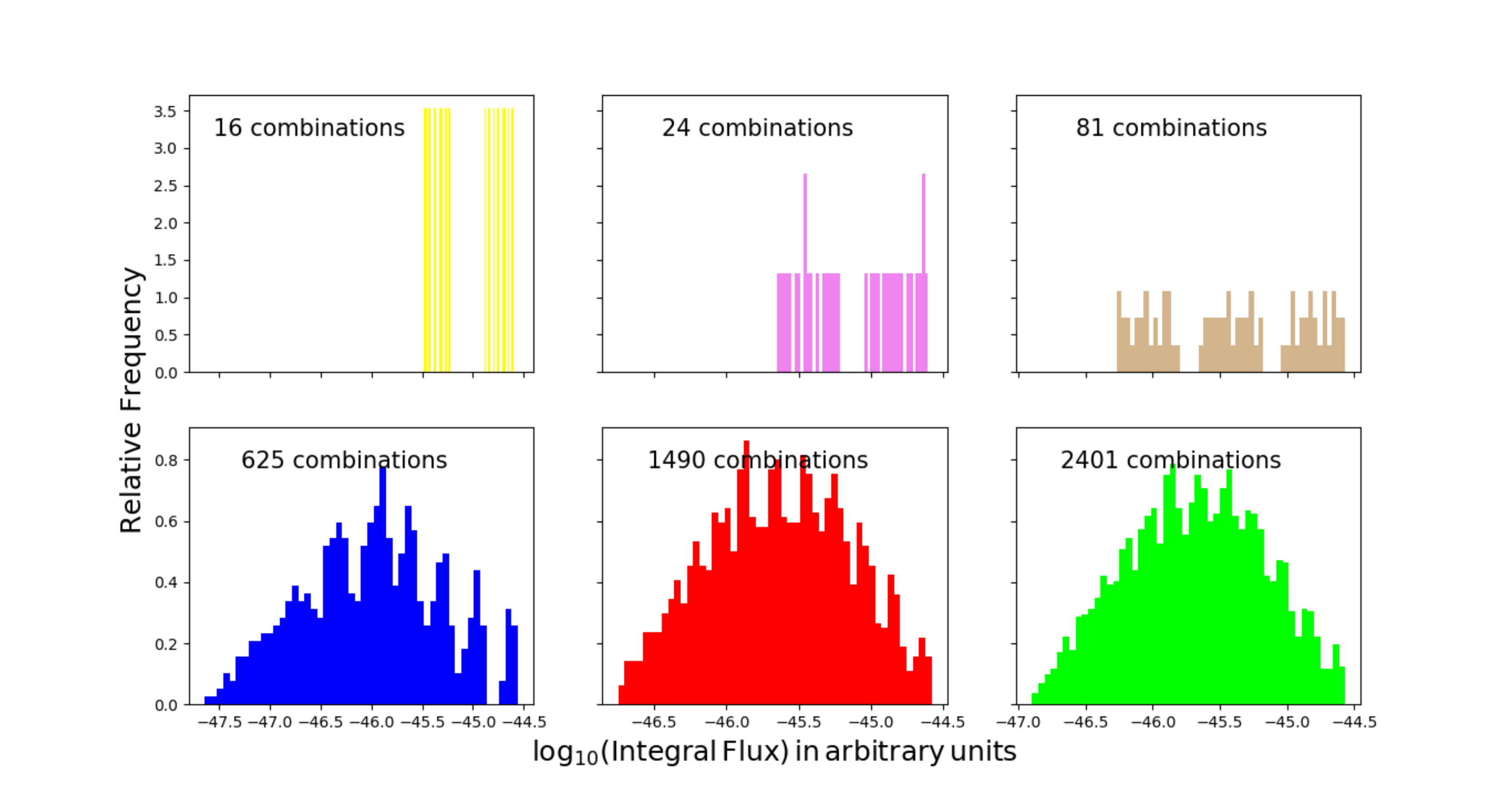}
\caption{Histograms of $\log_{10}$($F(E>0.72~{\rm TeV})$) in arbitrary units, indicating how we build up statistics for the flux distribution for a finer grid in parameter space. The $x$-axis shows the integral fluxes while the $y$-axis shows the normalised frequency. The label indicates the number of unique parameter combinations in each case.}\label{fig:build_up}
\bigbreak
\includegraphics[width=.8\textwidth]{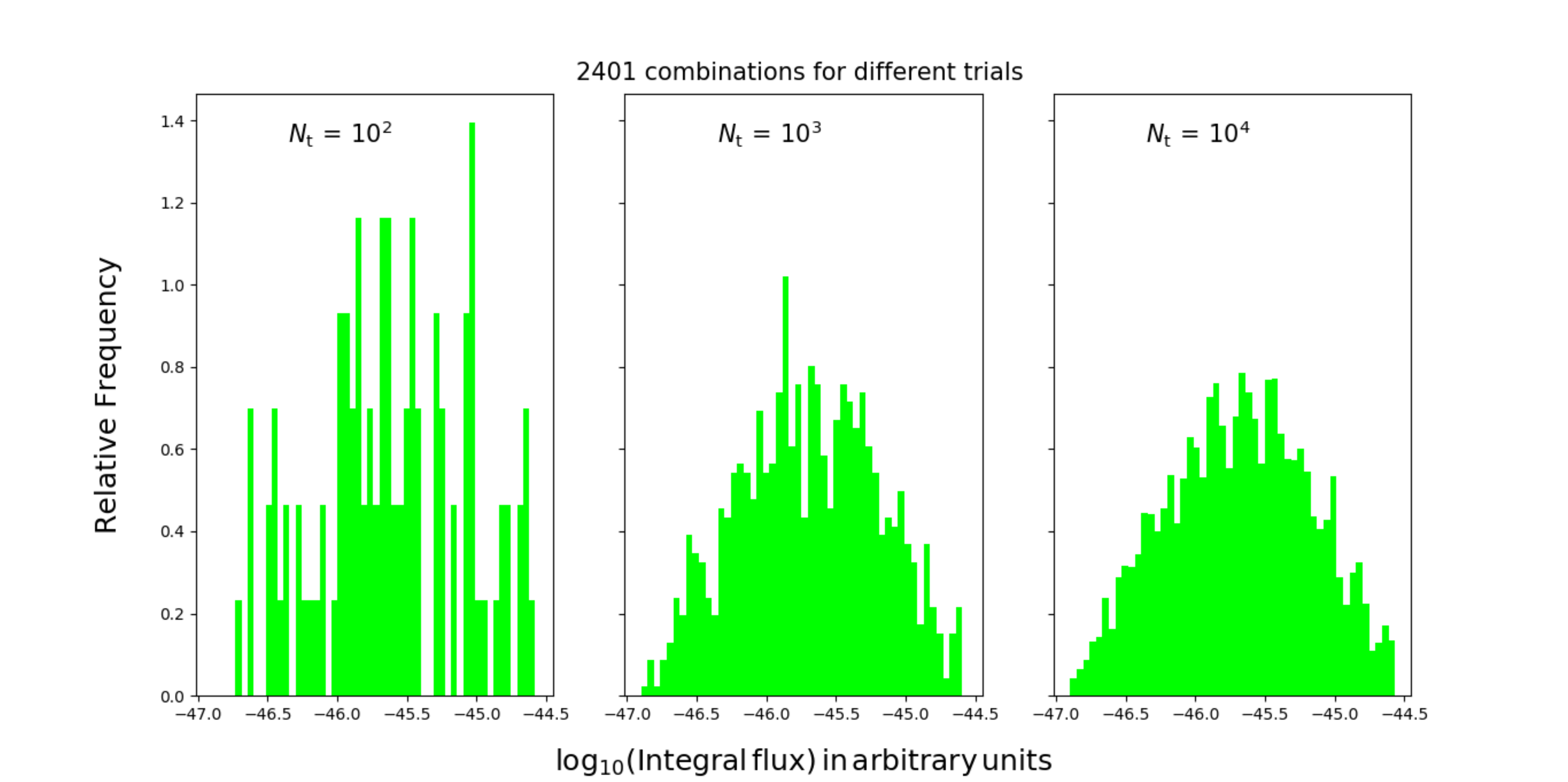}
\caption{Histograms of $\log_{10}$($F(E>0.72~{\rm TeV})$)} in arbitrary units, indicating  convergence for $N_{\rm t}$ getting close to number of combinations ($N_{\rm comb}$) and $N_{\rm t} > N_{\rm comb}$.\label{fig:two}
\end{figure*}

We first apply the model of \citet{Kopp2013} to the population of clusters to obtain the differential spectrum (Section~\ref{sec:model}) for different parameter combinations, initially setting $Q_{0} = 1$ and only using four free parameters ($B$, $\Gamma$, $N_*$, and $d$). We then calculate the integral flux using Equation~(\ref{eq:integral_flux}) and obtain a spread of fluxes, based on the range of values we picked for these four parameters.
To indicate the build-up of a smooth flux distribution, we use the four free parameters for illustration (for the full modeling, we use seven).
Figure~\ref{fig:build_up}
thus indicates how a distribution of $\log_{10}$ of the integral flux becomes smooth as we increase the resolution at which each of four free parameters is sampled. A finer grid 
%or more free parameters will
thus lead to a smoother 
%(and wider) 
distribution of fluxes. 

Second, we indicate convergence of the flux distribution with an increase of the number of trials, again for four free model parameters. That is, we calculate the integral flux for $N_{t}$ trials (each trial corresponding to a different parameter combination on a pre-defined grid), obtaining convergence as $N_{\rm t}$ gets close to or exceeds the number of unique  combinations ($N_{\rm comb}$) of the four free parameters. Thus, we reproduce the last panel of Figure~\ref{fig:build_up} to indicate convergence in Figure~\ref{fig:two}. We note that undersampling will not give a smooth flux distribution, but oversampling will lead to convergence.

\begin{figure*}
\centering
 \includegraphics[width=.7\textwidth]{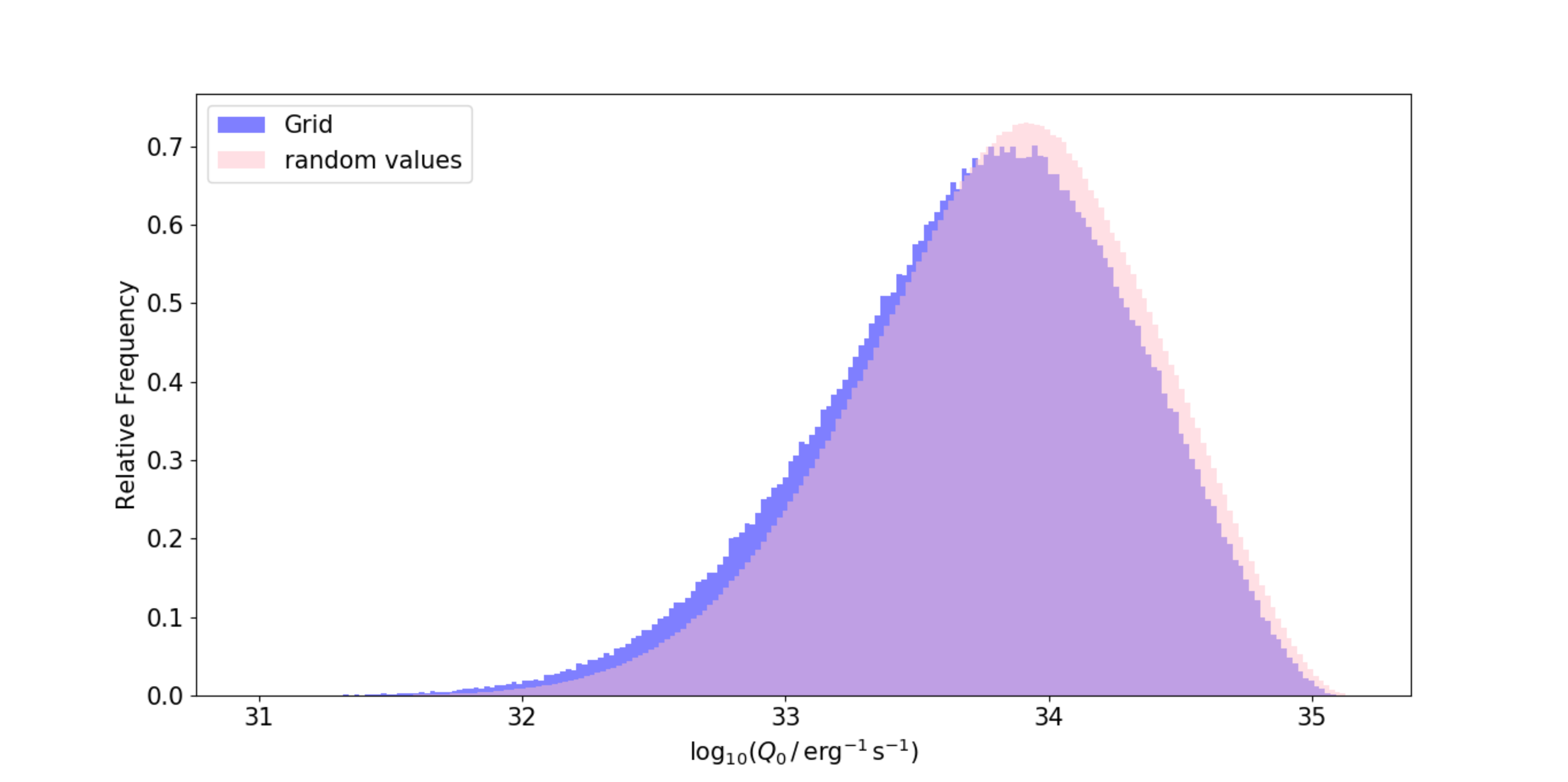}
\caption{Histogram of $\log_{10}(Q_{0})$ obtained via two different methods: sampling from a grid or randomly sampling for a set range for four free parameters. The pink histogram has a median $Q_{0,\mu} =7.0\times10^{33}\,{\rm erg^{-1}s^{-1}}$, $16^{\rm th}$ percentile of $1.8\times10^{33}\,{\rm erg^{-1}s^{-1}}$, and $84^{\rm th}$ percentile of $2.3\times10^{34}\,{\rm erg^{-1}s^{-1}}$ whilst the blue histogram has $Q_{0,\mu} = 5.9\times10^{33}\,{\rm erg^{-1}s^{-1}}$, $16^{\rm th}$ percentile of $1.4\times10^{33}\,{\rm erg^{-1}s^{-1}}$, and $84^{\rm th}$ percentile of $2.0\times10^{34}\,{\rm erg^{-1}s^{-1}}$}.
 \label{fig:four_parameters}
\end{figure*}

We thirdly check for the convergence of the source strength parameter $Q_{0}$ (see Figure~\ref{fig:four_parameters}) via two methods, using 47~Tucanae as an example. For the first method, we pick values of four relevant (different) parameters on a specified grid: $\eta \in [0.005, 0.08]$ in steps of 0.005, $\log_{10}\langle\dot{E}\rangle \in [33.7, 34.7]$ in steps of $0.05$, $N_{\rm MSP} \in [5, 150]$ in steps of 1, and $\Gamma \in [1.7, 2.9]$ in steps of $0.1$. For the second method, we randomise over the same four parameters $\eta$, $\log\dot{E}$, $N_{\rm MSP}$, and $\Gamma$, for fixed $E_{\rm e, min} = 0.1\,{\rm TeV}$ and $E_{\rm e, max} = 100\,{\rm TeV}$, within the parameter ranges given above (but not on a fixed grid). 
We conduct $N_{\rm t} = 10^{6}$ trials in both cases. The two different approaches yield very similar results, which also correspond to the values found by \citet{Kopp2013} as well as to the calculation by \citet{Venter2015}, with the two peaks in the histogram only slightly shifted with respect to each other. Thus, sampling on a grid vs.\ randomly picking parameter values does not make too much of a difference in terms of the distribution of $Q_0$.

\begin{figure*}
% \centering
% \includegraphics[width=0.8\textwidth]{Fig_3a.png}
%\caption{Histograms of $\log_{10}\rm(integral\,flux\,/\,cm^{-2}s^{-1})$ for 47~Tucanae, and for 7 free parameters.}\label{fig:47Tuc_integral_flux}
%\bigbreak
\includegraphics[width=0.9\textwidth]{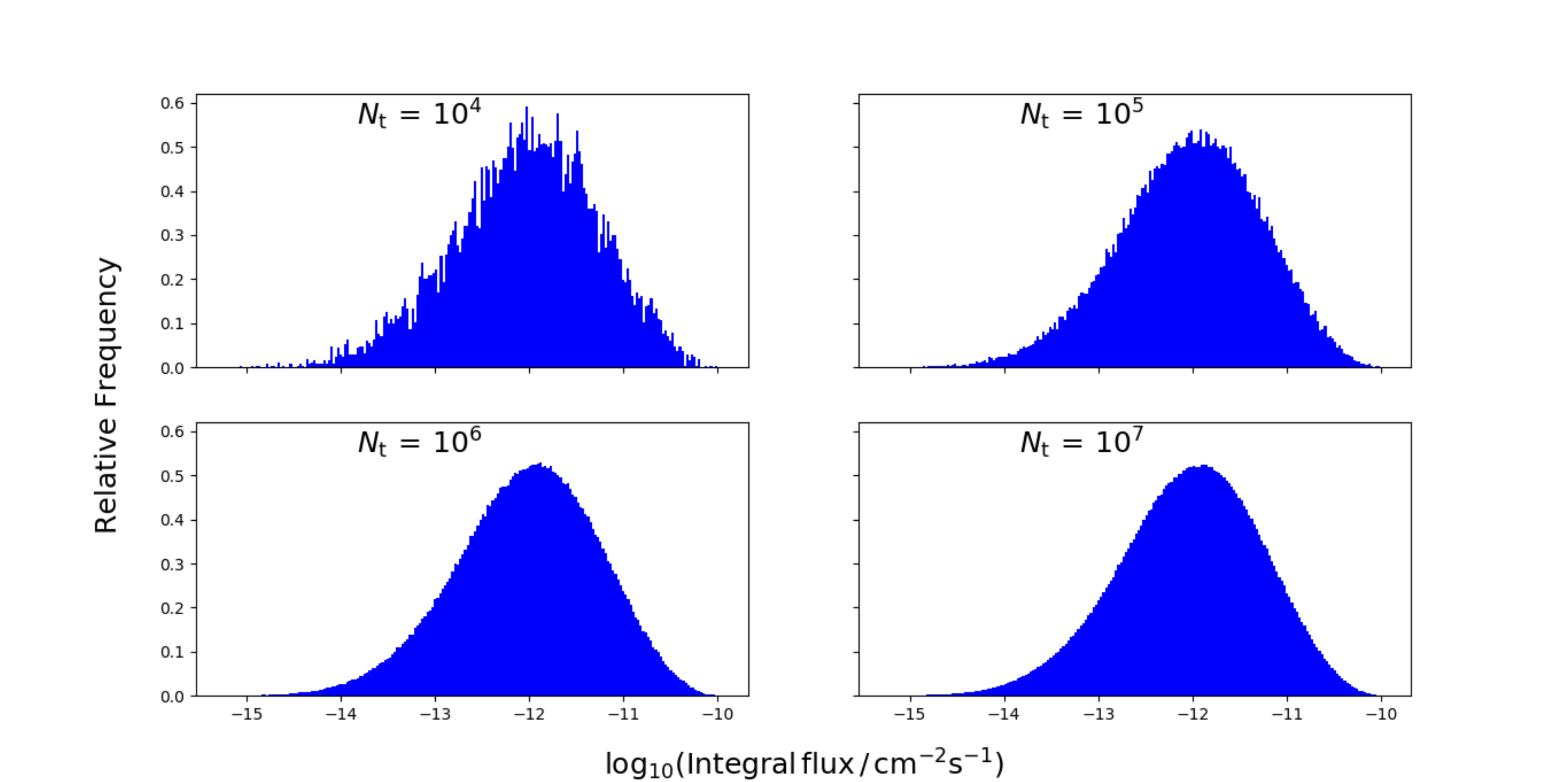}
\caption{Histograms of
 $\log_{10}(F(E>0.72~{\rm TeV})\,/\,{\rm cm^{-2}s^{-1})}$ for different $N_{\rm t}$ as indicated in each panel, for 47~Tucanae, and for 7 free parameters.}\label{fig:diff_Nt}
\end{figure*}

We lastly perform nested loops over the seven free model parameters to calculate $Q_{0}(\Gamma,\eta,\langle \dot{E}\rangle, N_{\rm MSP})\neq1$, and multiply the correct flux, pre-calculated for $Q_0 = 1$ and depending only on $B,\Gamma, d$, and $N_*$, by the actual value of $Q_0$ to obtain the final differential and integral fluxes in observational units 
(see Equation~[\ref{eq:normalisation}] and the final panel of Figure~\ref{fig:diff_Nt} for the case of 47~Tucanae).
The spread in flux reflects the uncertainty in model parameters. We use the observational threshold energy $E_{\rm th}$ associated with a particular cluster to calculate $F(>E_{\rm th})$. In Figure~\ref{fig:diff_Nt}, we  randomly sample integral fluxes for a different total number of trials $N_{\rm t}$ from the grid of pre-calculated fluxes to show convergence as $N_{\rm t}\rightarrow N_{\rm comb}$. 

\begin{figure*}
\centering
 \includegraphics[scale=0.5]{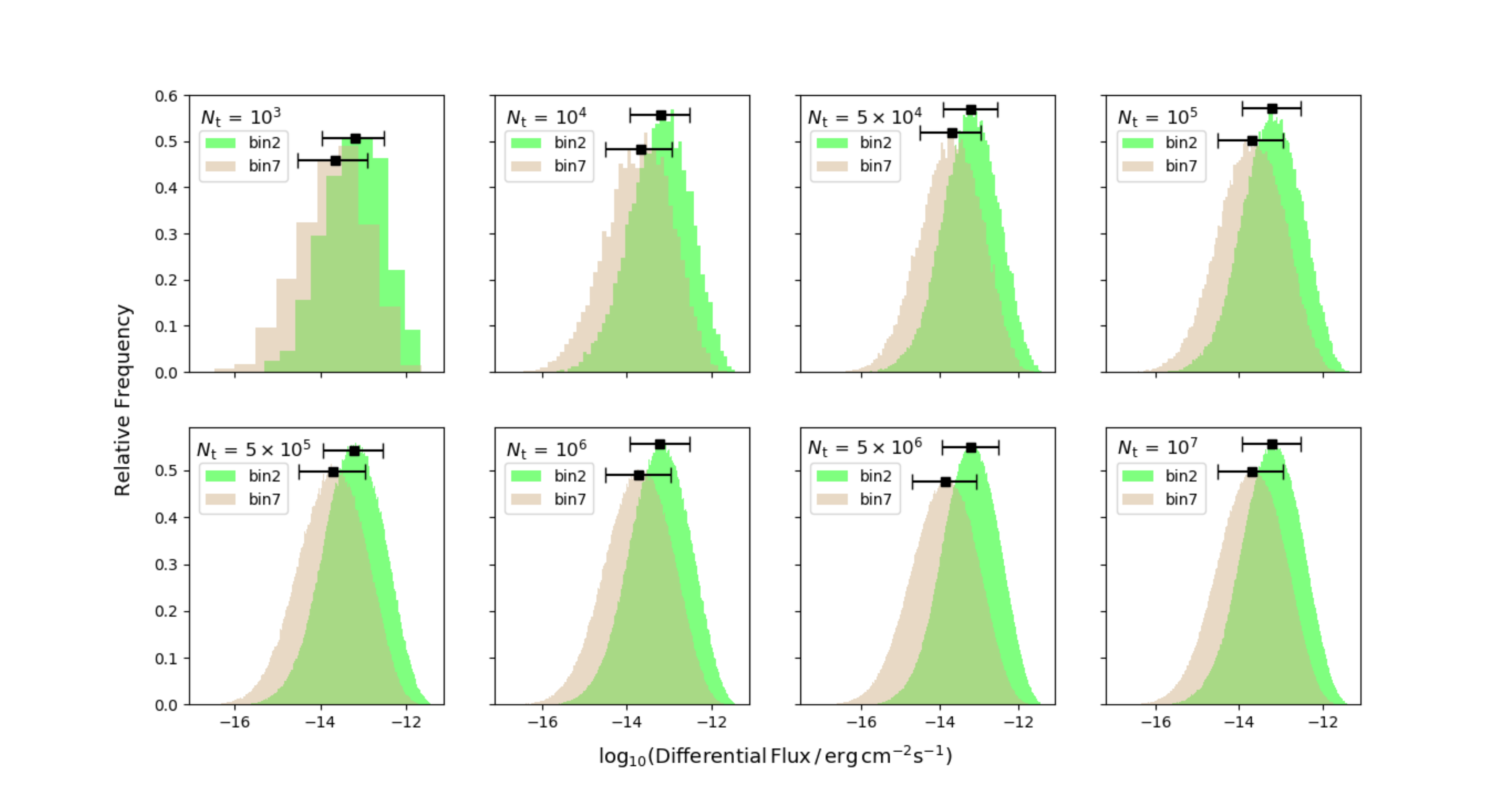}
 \caption{Distribution plots of $\log_{10}$ of the differential flux for MAGIC bin~2~($E = 62\,{\rm GeV}$) and bin~7~($E = 954\,{\rm GeV}$), with  the median and $1\sigma$ uncertainties indicated by the black square and error bars.
 The $x$-axis shows the $\log_{10}$ of the $\nu F_\nu$ flux while the $y$-axis shows the normalised frequency. The number of trials $N_{\rm t}$ in each case is indicated in each panel.}
 \label{fig:distribution}
\end{figure*}

\begin{figure*}
\centering
 \includegraphics[width=0.8\textwidth]{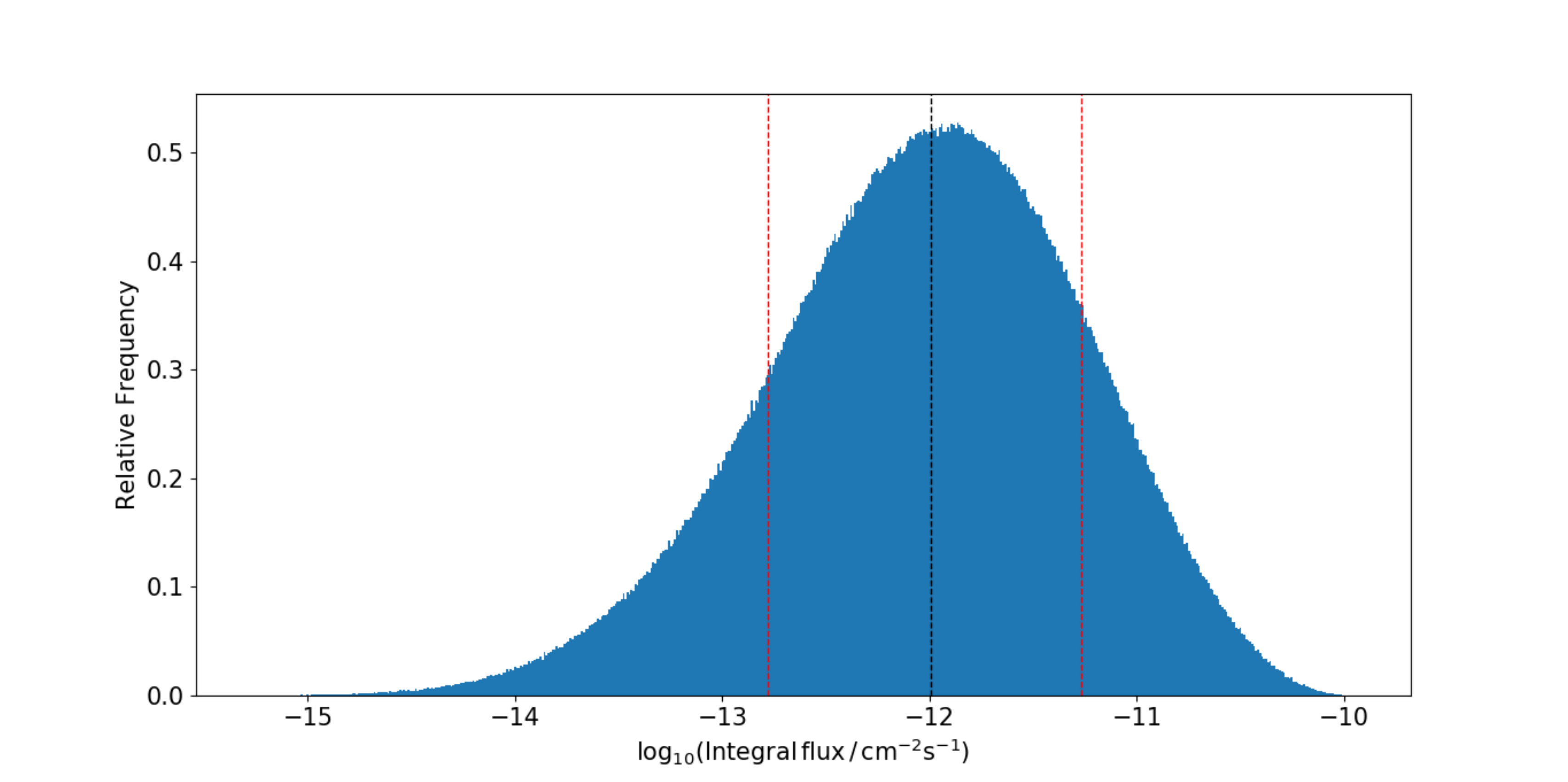}
\caption{Histogram of $\log_{10}(F(E>0.72~{\rm TeV})\,/\,{\rm cm^{-2}s^{-1})}$. The $x$-axis shows the $\log_{10}$ of integral flux while the $y$-axis shows the normalised frequency. The red dashed lines represent the $16^{\rm th}$ and $84^{\rm th}$ percentiles, and black dashed line shows the median of the integral flux distribution for 47~Tucanae.
 \label{fig:Fig4}}
\end{figure*}

\subsection{Differential Flux Predictions: M15 and $\omega$~Cen}
\label{MAGIC_method}
MAGIC has recently published differential flux upper limits for M15 (\citealt{Acciari2019}; see Section~\ref{M15_result} for details on the source and VHE data). We applied the model of \citet{Kopp2013} to M15 to obtain the \textit{differential} fluxes  for each combination of seven free model parameters (see Section~\ref{sec:model}). Next, we calculated the final differential flux as described in Section~\ref{sec:HESS_method}. 

We again use the method of \citet{Venter2008a} to estimate the uncertainty in the differential flux of M15 at 13 different observer-defined energy bins \citep{Acciari2019}. Specifically, we logarithmically interpolate model fluxes 
%(see Equation~\ref{eq:interpolated_flux}) 
to obtain predictions at the specified observational energies associated with the upper limits.
% We obtain and interpolate flux predictions using the model of \citet{Kopp2013} to get the prediction at the upper limit energies.
% The interpolated flux depends on the model input parameters. 
%\begin{equation}\label{eq:interpolated_flux}
% DF = e^{A}
%\end{equation}
%where $DF$ is the interpolated differential flux and 
%\begin{equation}
% A = (1-F)y_{1} + Fy_{2}
%\end{equation}
%with 
%\begin{equation}
% F = \frac{\log E_{\star} - \log E_{1}}{\log E_{2} - \log E_{1}}, 
%\end{equation}
%\begin{equation}
% y_{1} = \log(DF_{1}),
%\end{equation}
%and 
%\begin{equation}
% y_{2} = \log(DF_{2}).
%\end{equation}
Our parameter grid for M15 was chosen as follows: $B\in[1,\,3,\, 5,\, 7,\, 9]\,\mu G$, injection  spectral index $\Gamma\in[1.7,\, 2.0,\, 2.3,\, 2.6,\, 2.9]$,  $N_{*} \in [2.05,\, 2.90,\, 4.10,\, 5.80,\, 8.20]\times10^{5}$,  $d\in[5.20,\, 7.35,\, 10.4\, 14.7,\, 20.8]\,{\rm kpc}$, $\eta\in[0.003, 0.063]$ with steps of $0.005$, $\log_{10}\langle\dot{E}\rangle\in[33.7, 34.7]$ with steps of $0.05$, and the number of MSPs $N_{\rm MSP}\in[8, 24]$, in steps of $1$. In the case of $\omega$~Cen, we used the same $B$ and $\Gamma$ range as for M15, but $N_{*}\in[0.5,\, 0.7,\, 1.0,\, 1.4,\, 2.0]\times10^{6}$, $d\in [2.6,\, 3.7,\, 5.2,\, 7.6,\, 10.5]\,{\rm kpc}$, $\eta\in[0.005, 0.055]$, $\log_{10}\langle\dot{E}\rangle\in[33.9,34.9]$ in steps of $0.05$, and $N_{\rm MSP}\in[5, 299]$ in steps of 1. We conducted a maximum of $N_{\rm t}=10^7$ trials, with $N_{\rm comb} = 2.9\times10^6$ for M15 and $N_{\rm comb} = 7.4\times10^{6}$ for $\omega$~Cen, respectively (cf.\ Figure~\ref{fig:distribution}, where the value of $N_t$ is indicated in each panel).

\section{Results}
\subsection{The H.E.S.S.\ GC Population}
\begin{table*}
 \caption{H.E.S.S.\ flux upper limits from \citet{Abramowski2013} as well as model predictions corresponding to a first set of parameter combinations (associated with the green histograms in Figure~\ref{fig:Fig5a}). The columns in the Table are: the GC name; energy threshold of the analysis, defined as the location of the peak in the distribution of reconstructed photon energies; integral photon flux upper limits ($99\%$ confidence level following \citealt{Feldman1998}) assuming a power law with an index of 2.5 \citep{Abramowski2013} $F_{\rm UL}(E > E_{\rm th})$;  median of $\log_{10}$ of integral flux distribution $F_\mu$; ratio between the $\sigma_{16}$ and the median; ratio between $\sigma_{84}$ and the median; ratio between the median and the flux upper limit; geometric mean to characterise distributions with a significant variation across many orders of magnitude; ratio between standard deviation $\sigma$ and the median; and the ratio between the geometric mean and median. The bottom row represents results for a stacking scenario, i.e., $F_{\rm UL}(E > E_{\rm th})$ represents the live-time-weighted flux of 15 GCs, while the modelling results (cf.\ Eq.~\ref{eq:Fw}) are for the same 15 GCs, and weighted similarly.} 
 %It is useful in characterizing distributions where you expect significant variation across many orders of magnitude.}
\label{table:1}
\begin{center} 
\begin{tabular}{|c | c | c | p{1cm} | c | c | c | c | c | c|}          
\hline                      
GC name & \vtop{\hbox{\strut$E_{\rm th}$}\hbox{\strut(${\rm TeV}$)}} & \vtop{\hbox{\strut$F_{\rm UL}$($E > E_{\rm th}$)}\hbox{\strut$\times10^{-13}$}\hbox{\strut($\rm ph\,cm^{-2} s^{-1}$)}} & \vtop{\hbox{\strut $F_{\mu}$} \hbox{\strut$\times10^{-13}$}\hbox{\strut($\rm ph\,cm^{-2} s^{-1}$)}} & \vtop{\hbox{\strut$\sigma_{16}/F_\mu$}\hbox{\strut($\%$)}} & \vtop{\hbox{\strut$\sigma_{84}/F_\mu$}\hbox{\strut($\%$)}} & ${F_{\mu}}/F_{\rm UL}$ & \vtop{\hbox{\strut ${\bar{F}_{G}}$} \hbox{\strut$\times10^{-13}$}\hbox{\strut(${\rm ph\,cm^{-2}\,s^{-1}}$)}} & \vtop{\hbox{\strut$\sigma/F_\mu$}\hbox{\strut($\%$)}} &  ${\bar{F}_{G}/F_\mu}$ \\
\hline
	 NGC 104   & 0.72 & 19 & 10.0 & 84 & 427 & 0.53 & 9.4 & 611 & 0.92  \\
	 NGC 6388  & 0.28 & 15 & 5.9  & 85 & 461 & 0.39 & 5.5 & 706 & 0.93  \\
     NGC 7078  & 0.40 & 7.2& 1.7  & 84 & 465 & 0.24 & 1.5 & 673 & 0.94  \\
 	 Terzan 6  & 0.28 & 21 & 6.9  & 86 & 518 & 0.33 & 6.3 & 890 & 0.91  \\
 	 Terzan 10 & 0.23 & 29 & 11.0 & 87 & 530 & 0.38 & 10.0& 903 & 0.91  \\
 	 NGC 6715  & 0.19 & 9.3& 1.2  & 85 & 478 & 0.13 & 1.1 & 722 & 0.92  \\ 
	 NGC 362   & 0.59 & 24 & 1.5  & 88 & 572 & 0.06 & 1.4 & 1081& 0.93  \\
	 Pal 6     & 0.23 & 12 & 11.0 & 87 & 534 & 0.92 & 9.9 & 919 & 0.90  \\
	 NGC 6256  & 0.23 & 32 & 3.2  & 87 & 534 & 0.10 & 2.9 & 941 & 0.91  \\  
	 Djorg 2   & 0.28 & 8.4& 10.0 & 86 & 508 & 1.19 & 9.2 & 831 & 0.92  \\
	 NGC 6749  & 0.19 & 14 & 6.9  & 87 & 534 & 0.49 & 6.3 & 913 & 0.91  \\
	 NGC 6144  & 0.23 & 14 & 4.3  & 87 & 554 & 0.31 & 3.9 & 972 & 0.91  \\    
	 NGC 288   & 0.16 & 5.3& 6.1 & 87 & 553 & 1.15 & 5.6 & 945 & 0.92  \\
	 HP 1      & 0.23 & 15 & 5.4  & 87 & 539 & 0.36 & 5.0 & 925 & 0.93  \\  
	 Terzan 9  & 0.33 & 45 & 2.8  & 89 & 644 & 0.06 & 2.6 & 1341& 0.93  \\ 
 \hline
Stacking analysis & 0.23 & 3.3 & 25  & 50 & 119 & 7.6  & 26.0& 135 & 1.04  \\
 \hline
\end{tabular}
\end{center}
\end{table*}

\begin{figure*}
\centering
 \includegraphics[width=0.9\textwidth]{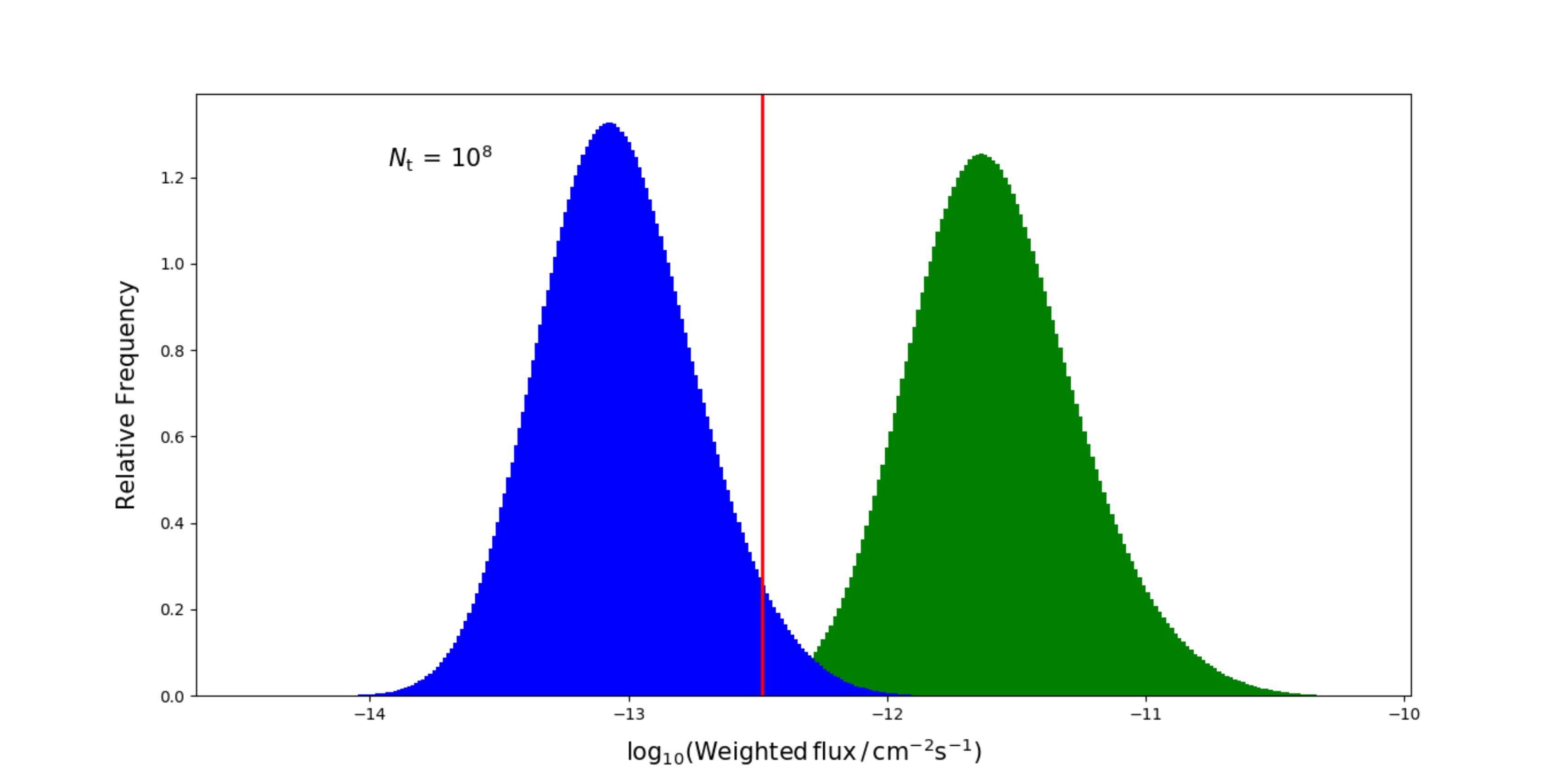}
\caption{Distribution of $\log_{10}(F_{\rm w}(E>E_{\rm th})\,/\,{\rm cm^{-2}s^{-1})}$ for all GCs. The $x$-axis shows the $\log_{10}$ of total weighted flux while the $y$-axis shows the normalised frequency. The green histograms are associated with the first parameter combination, while the blue ones are for the second parameter combination. The red line represents the stacked upper limit as noted in Table~\ref{table:1}. We used $N_{\rm t}=10^8$.}
 \label{fig:Fig5a}
\end{figure*}

\begin{figure*}
\centering
 \includegraphics[scale=0.5]{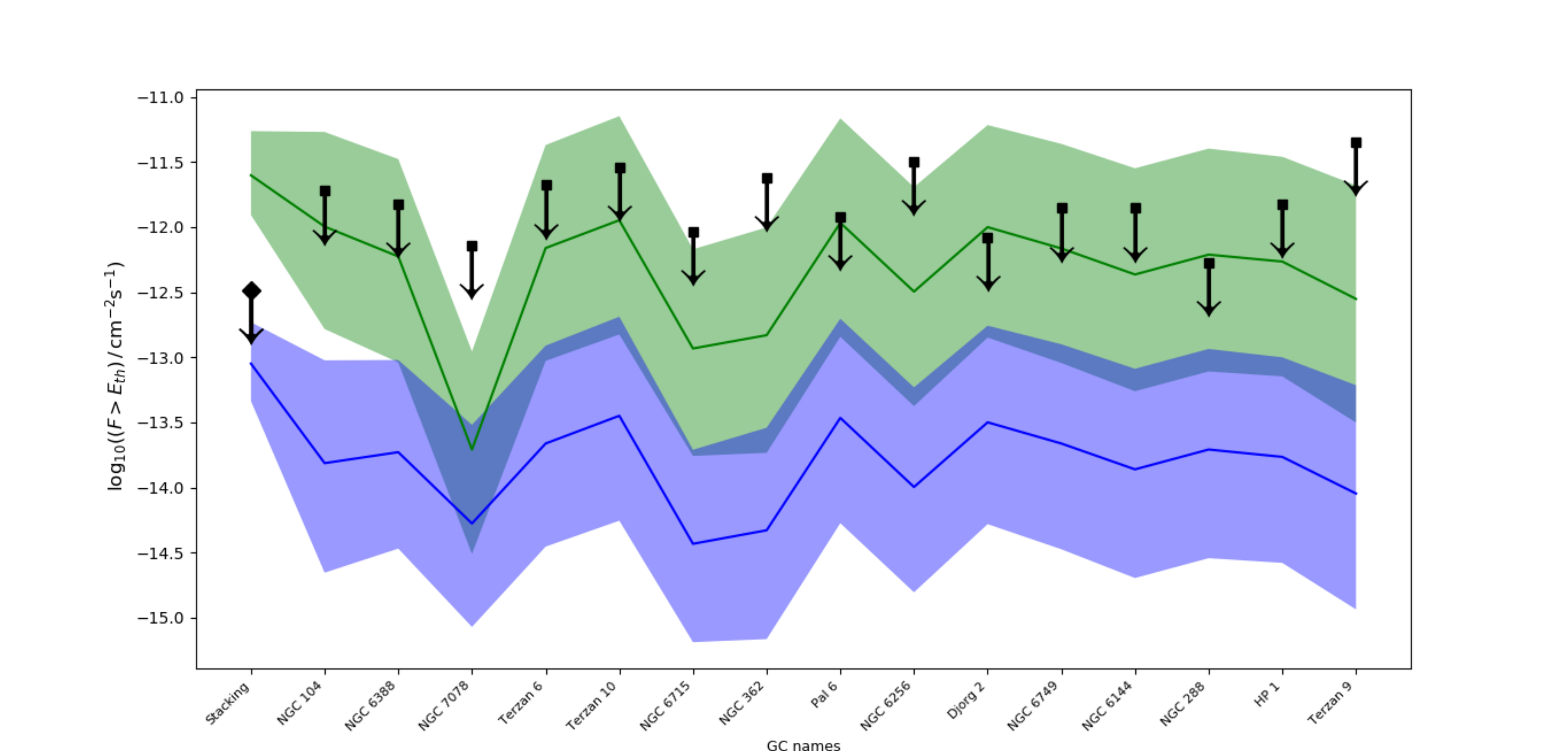}
 \caption{Flux upper limits (black diamond and squares with arrows) on the observed $\gamma$-ray flux from the population of GCs, as well as the predicted flux medians plus $1\sigma$ uncertainties (green and blue) for each of the two parameter combinations. The first entry indicates the stacked (live-time-weighted) flux with 1$\sigma$ uncertainties, as indicated in Table~\ref{table:1}.}
 \label{fig:UL}
\end{figure*}

\begin{table*}
 \caption{H.E.S.S.\ flux upper limits and model predictions for the second parameter combination, associated with the blue histograms in Figure~\ref{fig:Fig5a}.}
\label{table:2}
\centering 
\begin{tabular}{|c | c | c | p{1cm} | c | c | c | c | c | c|}          
\hline                      
GC name & \vtop{\hbox{\strut$E_{\rm th}$}\hbox{\strut(${\rm TeV}$)}} & \vtop{\hbox{\strut$F_{\rm UL}$($E > E_{\rm th}$)}\hbox{\strut $\times10^{-13}$}\hbox{\strut($\rm ph\,cm^{-2} s^{-1}$)}} & \vtop{\hbox{\strut $F_{\mu}$} \hbox{\strut$\times10^{-13}$}\hbox{\strut($\rm ph\,cm^{-2} s^{-1}$)}} & \vtop{\hbox{\strut$\sigma_{16}/F_{\mu}$}\hbox{\strut($\%$)}} & \vtop{\hbox{\strut$\sigma_{84}/F_{\mu}$}\hbox{\strut($\%$)}} & $F_{\mu}/F_{\rm UL}$ & \vtop{\hbox{\strut ${\bar{F}_{G}}$} \hbox{\strut$\times10^{-13}$}\hbox{\strut(${\rm ph\,cm^{-2}\,s^{-1}}$)}} & \vtop{\hbox{\strut$\sigma/F_{\mu}$}\hbox{\strut($\%$)}} & ${\bar{F}_{G}/F_{\mu}}$   \\
\hline
	 NGC 104   & 0.72 & 19 & 0.154 & 86 & 614 & 0.0081 & 0.145 & 929 & 0.94 \\
	 NGC 6388  & 0.28 & 15 & 0.187 & 82 & 410 & 0.0125 & 0.179 & 610 & 0.96  \\
	 NGC 7078  & 0.40 & 7.2  & 0.053 & 84 & 573 & 0.0074 & 0.051 & 737 & 0.95 \\
 	 Terzan 6  & 0.28 & 21 & 0.218 & 84 & 463 & 0.0104 & 0.208 & 766 & 0.92  \\
 	 Terzan 10 & 0.23 & 29 & 0.356 & 84 & 475 & 0.0123 & 0.337 & 778 & 0.95  \\
 	 NGC 6715  & 0.19 & 9.3  & 0.037& 82 & 429  & 0.0040 & 0.035 & 622 & 0.96 \\
	 NGC 362   & 0.59 & 24 & 0.047 & 85 & 513 & 0.0020 & 0.044 & 923 & 0.94 \\  
	 Pal 6     & 0.23 & 12 & 0.343 & 85 & 578 & 0.0286 & 0.324 & 787 & 0.94 \\ 
	 NGC 6256  & 0.23 & 32 & 0.101 & 84 & 482 & 0.0032 & 0.096 & 810 & 0.95 \\
	 Djorg 2   & 0.28 & 8.4  & 0.317 & 84 & 451 & 0.0377 & 0.302 & 713 & 0.95 \\
	 NGC 6749  & 0.19 & 14 & 0.218 & 84 & 577 & 0.0156 & 0.205 & 780 & 0.94 \\
	 NGC 6144  & 0.23 & 14 & 0.138 & 86 & 492 & 0.0097 & 0.129 & 819 & 0.93  \\ 
	 NGC 288   & 0.16 & 5.3  & 0.196 & 85 & 492 & 0.0370 & 0.183 & 806 & 0.93 \\
	 HP 1      & 0.23 & 15 & 0.172 & 85 & 481 & 0.0115 & 0.162 & 797 & 0.95 \\ 
	 Terzan 9  & 0.33 & 45 & 0.090 & 87 & 580 & 0.0020 & 0.084 & 1147& 0.93 \\ 
 \hline
Stacking analysis&0.23& 3.3  & 0.895& 48 & 108  & 0.271  & 0.923 & 113 & 1.03 \\  
 \hline
\end{tabular}
\end{table*}

To obtain a sample of GCs suited for the study by H.E.S.S.,\
\citet{Abramowski2013} selected a population of clusters based on the GC catalog of \citet{Harris2010} from which all the basic parameters such as position, $d$, $r_{\rm c}$, and central luminosity density $\rho_{0}$ have been taken.  \citet{Abramowski2013} furthermore applied different cuts on the target and observation run lists:\\
1. GC $|{\rm Galactic\,latitude}|\geq 1.0^\circ$. This cut was done to conservatively exclude GCs towards the direction of the Galactic Plane to avoid faint emission from unresolved sources, diffuse emission, and also chance coincidences with unrelated sources.\\
2. Standard quality selection criteria were applied, with the telescopes pointed $<2.0^\circ$ offset from the GC position.\\
3. A selected GC had to have at least $20$ available runs passing the cut describe in point $2.$ to ensure a reasonably long exposure on these potentially faint sources.

In order to compare our model predictions with the H.E.S.S.\ flux upper limits $F_{\rm UL}$, we have to characterise the mean predicted flux plus its uncertainty\footnote{We note that it is common in certain statistical contexts to use the $2^{\rm nd}$ quartile (median) as an indication of an average value, and the $1^{\rm st}$ and $3^{\rm rd}$ quartiles ($25^{\rm th}$ and $75^{\rm th}$ percentiles) as indications of uncertainty or the spread of values around the median, as is done in box plots. However, we opt to use the median and the $16^{\rm th}$ and $84^{\rm th}$ percentiles to estimate the average and uncertainty in model flux (in analogy to what is typical for a normal distribution of values), as is commonly done in certain physics contexts, given the fact that our flux distributions appear to be approximately log-normal. Our main conclusion, that there are large uncertainties on the median model fluxes due to uncertainty on model parameters, is not affected by this choice of convention.} First, we obtain the median flux $F_\mu$ ($50^{\rm th}$ percentile) for each cluster, as well as the $16^{\rm th}$ and $84^{\rm th}$ percentile, as shown in Figure~\ref{fig:Fig4} and listed in Table~\ref{table:1}. 
We calculate $\sigma_{16} = \mu - F_{16}$ and  $\sigma_{84} = F_{84} - \mu$ and also express them as percentages: $100\sigma_{16}/\mu$ and $100\sigma_{84}/\mu$, with $F_{16}$ and $F_{84}$ being the $16^{\rm th}$ and $84^{\rm th}$ percentiles.
We also calculate the geometric mean of the flux $\bar{F}_{\rm G}$ (since we want to be insensitive to outliers), as well as the standard deviation $\sigma$ of this flux, noting that the geometric mean is equal to the antilog of the arithmetic mean of the log of fluxes. We finally list the ratios $F_\mu/F_{\rm UL}$ as well as $\bar{F}_{\rm G}/F_\mu$, indicating that the median is usually below the upper limit, except for Djorg 2 and the stacking upper limit, and that $F_\mu \approx \bar{F}_{\rm G}$.
%That is, if we consider $N$ values of $x_j$, $j = 1,2\dots,N$. The geometric mean is defined as
%$$\langle x \rangle_{\rm G} = (x_1x_2\dots x_N)^{1/N}$$. \\
%Considering the arithmetic mean of the log of the $x$ values:
%$$\langle x \rangle_{\rm A,log} = \frac{1}{N}\sum_{i=1}^N \log(x_i) =  \frac{1}{N} \log(\Pi_{i=1}^Nx_i) =\log[(\Pi_{i=1}^Nx_i)^{1/N}] $$
%Therefore
%$$10^{\langle x \rangle_{\rm A,log}} = \langle x \rangle_{\rm G}.$$

%We also obtain the median, $16^{\rm th}$ percentile, $50^{\rm th}$ percentile, and $84^{\rm th}$ percentile noting that the median ($\mu$) is the same as the $50^{\rm th}$ percentile and that they are equal for both linear and log of fluxes. 

In Figure~\ref{fig:Fig5a} we show the distribution of the $\log_{10}$ of weighted (stacked) flux for all GCs. We plot the weighted flux as a function of $N_{\rm t}$ for all 15 GCs. We use  $E_{\rm th} = 0.23$~TeV and calculate the weighted flux as
\begin{equation}
F_{\rm w} = \sum_{i=1}^{15}\frac{\rm \tau_i}{\tau_{\rm tot}}\times F_i(>0.23~{\rm TeV}),\label{eq:Fw}   
\end{equation}
where $i$ represents a specific GC, $\tau_i$ is the live time (acceptance-corrected observational time passing quality cuts) for each GC, and $\tau_{\rm tot}$ is the total live time. The median and geometric mean of this quantity are listed in the bottom row of Table~\ref{table:1}, in column 4 and 8, respectively, for comparison to the stacked upper limit published by H.E.S.S.\ as listed in column 3. The first set (range) of parameter combinations used for the green histograms in Figure~\ref{fig:Fig5a} is as follows: $d_{\rm cluster} \in [\frac{d}{2}, \frac{d}{\sqrt{2}}, d, \sqrt{2}d, 2d]$, $N_{*} \in [\frac{N}{2}, \frac{N_*}{\sqrt{2}}, N_{*}, \sqrt{2}N_{*}, 2N_{*}]$ (see \citealt{Venter2015} for values of $d$ and $N_{*}$). We used $B \in [1,9]$ in steps of 2, $\Gamma \in [1.7, 2.9]$ with steps of 0.3, $\eta \in [0.005, 0.08]$ in steps of 0.0075, $\log_{10}\langle\dot{E}\rangle \in [33.7, 34.7]$ in steps of 0.1, and $N_{\rm MSP} \in [5,150]$ in steps of 10. Although somewhat arbitrary, these were chosen to be an example set of reasonable (best-guess) parameter values. However, we see from Table~\ref{table:1} that for this initial choice, the median stacked flux exceeds the upper limit by a factor of 7.6. We therefore considered a second parameter set, associated with the blue histograms of weighted flux for 15 GCs in Figure~\ref{fig:Fig5a}: $\eta \in [0.003, 0.03]$ in steps of 0.003, $\log_{10}\langle\dot{E}\rangle \in [33, 34]$ in steps of 0.1, and $N_{\rm MSP} \in [5,50]$ in steps of 5. Again, this choice might seem arbitrary, but given the freedom in model parameters, there is no unique way of reducing the GC flux. Thus, we reduced the three (degenerate) main parameters determining the source strength $Q_0$, i.e., $\eta$, $\dot{E}$, and $N_{\rm MSP}$ by a factor of a $\sim 3$ to illustrate that by lowering $Q_0$, the flux upper limits may be satisfied. While we do not claim this second set as a final answer, we deem these ranges to still be realistic, given the uncertainty in their intrinsic values. The fluxes obtained for the second choice of parameter grid are indicated in Table~\ref{table:2}. In this case, the predicted median flux is a factor of $\sim4$ below the H.E.S.S.\ upper limit (indicated by the red solid line in Figure~\ref{fig:Fig5a}).  
This is one example (i.e., changing some parameters so as to yield a smaller $Q_0$) of how one can use the H.E.S.S.\ data to constrain the MSP population's characteristics \citep[see also][]{Ndiyavala2019}, although there may be some degeneracy with other model parameters such as $\kappa$, $\Gamma$, $d$ and to a lesser extent $B$ (this will impact the SR component more) and $N_*$, although the latter is relatively well constrained.

%Figure~\ref{fig:Fig5a} shows our weighted fluxes for the population of 15 GCs.  The green histograms is for the first set of parameters as discussed in Section~\ref{sec:HESS_method} and the results is shown in Table~\ref{table:1}. These particular combinations of parameters violates the H.E.S.S.\ stacked upper limits. However, there are indeed parameter combinations for which the total weighted integral flux is below the H.E.S.S.\ upper limits (e.g., the blue histogram with Table~\ref{table:2} results) given the assumptions for the free parameter ranges. 
Finally, we graphically summarise the fluxes we obtained for both the stacked as well as single-GC cases in Figure~\ref{fig:UL}. The black diamond and squares and arrows correspond to the observational upper limits, while the green and blue squares and error bars are for the median fluxes and $1\sigma$ uncertainty intervals, for the two different parameter combinations as discussed above.
We note that there are indeed parameter combinations that yield fluxes below all of these observational upper limits. One example is afforded by combinations of degenerate parameters yielding a lower $Q_0$; there may also be other parameter combinations that can lower the flux, as indicated before. Further constraints derived from future observations may aid in breaking such degeneracies.

\subsection{M15} \label{M15_result}
The northern GC M15 (also known as NGC 7078) is one of the densest\footnote{https://www.nasa.gov/feature/goddard/2017/messier-15} globular clusters ever discovered in the Milky Way Galaxy. It has undergone a core collapse contraction and thus belongs to the class of core-collapsed brightest GCs \citep{Acciari2019}. M15 hosts at least eight MSPs detected so far \citep{Freire2015}. It is located at a distance of $d = 10.4\,{\rm kpc}$ \citep{Harris1996} at RA (J2000) $21^{\rm h}29^{\rm m}58^{\rm s}.33$ and Dec $+12^\circ10^\prime01^{\prime\prime}.2$ (Galactic coordinates: $l = 65.01^\circ, b = -27.31^\circ$) and exhibits $r_{\rm c} = 0^\prime.14$ and a half-light radius $r_{\rm h} = 1^\prime.00$ \citep{Harris1996}.

\citet{Acciari2019} observed M15 with the MAGIC telescopes, since it is the only GC discovered by \emph{Fermi} LAT at GeV $\gamma$-ray energies that they could observe at low zenith angles. Its $\gamma$-ray luminosity measured above $0.1\,{\rm GeV}$ is $L_{\gamma}^{\rm M15} = \left(5.26_{-1.16}^{+1.31}\right)\times10^{34}\,{\rm erg\,s^{-1}}$ and the spectrum is similar to the power-law type with spectral index 2.84 $\pm$ 0.18 when observed up to $\sim5$ GeV \citep{Zhang2016}. 
The estimated number of MSPs in M15 is $37_{-8}^{+9}$, for a conversion efficiency of the MSP's spin-down power into GeV $\gamma$-rays of $\eta_{\gamma}\approx 0.08$ and an average luminosity of $\langle L_{\gamma}^{\rm MSP}\rangle = 1.44\times10^{33}\,{\rm erg\,s^{-1}}$ per MSP. However, no detection from the direction of M15 has been made in the TeV band \citep{Acciari2019}. These authors therefore followed \citet{Rolke2005} using a $95\%$ confidence level and assuming a $30\%$ total systematic uncertainty on the collection area and spectral index of $-2.6$ to calculate the upper limits on the VHE flux from the source. They found an upper limit of $F(E>0.3~{\rm TeV})=3.2\times10^{-13}\,{\rm cm^{-2}s^{-1}}$, which several times more stringent than  the $F(E>0.44~{\rm TeV})=7.2\times10^{-13}\,{\rm cm^{-2}s^{-1}}$ published earlier by \citet{Abramowski2013}.

We create a distribution of differential fluxes (see Figure~\ref{fig:distribution} for energy bins 2 and 7, centred on energies of $E=62$~GeV and $E=955$~GeV, respectively) and calculate the median and $1\sigma$ uncertainties of the $\log_{10}$ of these fluxes by finding the $16^{\rm th}, 50^{\rm th}$ and $84^{\rm th}$ percentiles. %Figure~\ref{fig:distribution} shows the spread of interpolated fluxes with median and $1\sigma$ errors for MAGIC energy bin~2~($E = 1.79\,{\rm GeV}$) and bin~7~($E = 2.98\,{\rm GeV}$. 
As before, the fluxes were obtained by randomising over seven free parameters many times (see the different $N_{\rm t}$ as indicated in each panel) to study the convergence.
 
\begin{figure}
\centering
 \includegraphics[height=5.5cm, width=8.5cm]{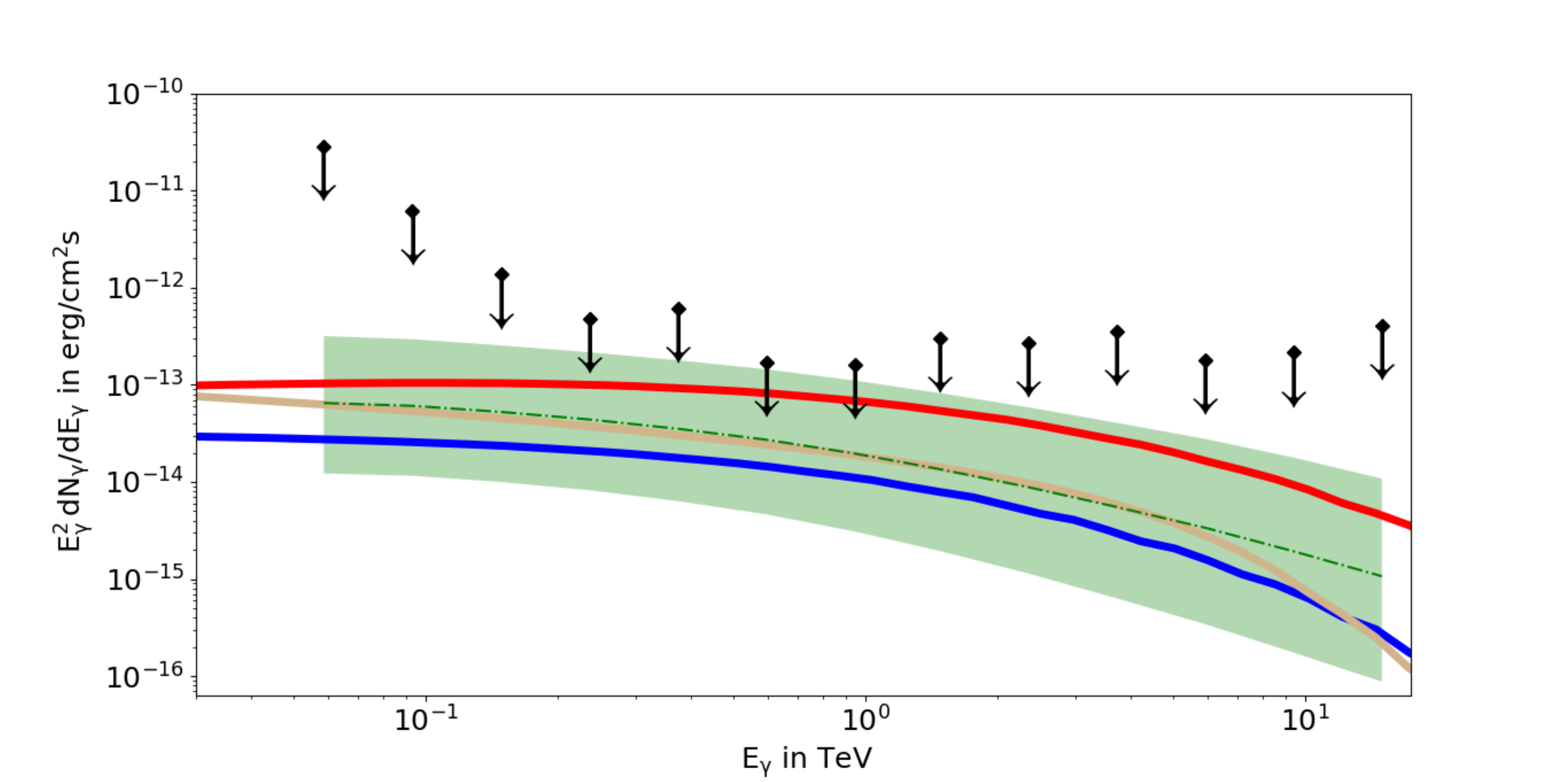}
 \caption{Differential flux upper limits (black diamonds with arrows) on the $\gamma$-ray flux from M15 plus a few typical model spectra (solid lines) for typical parameter combinations, as well as predicted medians and $1\sigma$ error bars (green error band). For the example fits, we used lepton energies of $E_{\rm min} = 0.01\,{\rm TeV}$, $E_{\rm max} = 100\,{\rm TeV}$, assumed Bohm diffusion, $B = 5\mu G, \Gamma = 1.9$, and $Q_{0} = 1.81\times10^{33}\,{\rm erg^{-1}\,s^{-1}}$ (red line); $E_{\rm min} = 0.01\,{\rm TeV}, E_{\rm max} = 30\,{\rm TeV}, k_{0} = 7\times10^{-5}\,{\rm kpc^{2}/Myr}, B = 3\mu G, \Gamma = 2.1$, and $Q_{0} = 4.75\times10^{32}\,{\rm erg^{-1}\,s^{-1}}$ (blue line) and $E_{\rm min} = 0.01\,{\rm TeV}, E_{\rm max} = 30\,{\rm TeV}, k_{0} = 1.0\times10^{-3}\,{\rm kpc^{2}/Myr}, B = 5\mu G, \Gamma = 1.9$, and $Q_{0} = 1.81\times10^{33}\,{\rm erg^{-1}\,s^{-1}}$ (golden line).}
 \label{fig:M15}
\end{figure}
Figure~\ref{fig:M15} shows the upper limits from 165 hrs of MAGIC observations on the differential flux, the predicted spectrum for a few different combinations of model parameters, and the median points with $1\sigma$ error. 
%We also show typical differential spectra as compared by the model. 
We note that the differential data are much more constraining than the integral data since there are 13 points of comparison instead of 1. For each bin, we are able to tell whether the model exceeds the flux at that energy bin or not, thus constraining the spectral shape. For our choice of parameters, we satisfy the MAGIC flux upper limits.

\subsection{$\omega$~Cen}
$\omega$ Centauri (NGC 5139), which is located at RA~$201.7^\circ$, Dec~$-47.56^\circ$,
stands out among all the known GCs in our Galaxy. Its origin and history are not fully understood yet, but its properties have led to suggestions that it may be the remnant core of a dwarf galaxy that has been disrupted (e.g., \citealt{Bekki2003}). $\omega$~Cen is the most massive, complex, brightest, and has the largest core and half-light radius of all GCs in the Galaxy \citep{Harris1996}. It is thought to have been the nuclear star cluster of either the Sequoia or Gaia-Enceladus Galaxy discussed in \citet{Myeong2019}. The cluster also shows characteristic features such as its incredible multiplicity in stellar populations (e.g., \citealt{Pancino2000, Bellini2017}) and its broad metallicity distribution (e.g., \citealt{Freeman1975, Magurno2019}). $\omega$~Cen is furthermore suspected to harbour a black hole with total mass $\sim10^{5}M_{\odot}$ at its centre, which makes this GC an interesting candidate to model for possible future CTA observations \citep{Zocchi2019}. 

Observations by \textit{Fermi} LAT has indicated that $\omega$~Cen's spectrum was reminiscent of a typical MSP spectrum. However, it was a mystery that no radio or X-ray MSPs had been detected \citep{Abdo2010} at that time. The Parkes Radio Telescope has now detected five radio MSPs in this cluster, four of which are isolated and a fifth one is in an eclipsing binary system. This confirms the scenario that the cumulative MSP $\gamma$-ray emission may explain the \textit{Fermi} detection of this cluster. However, no pulsations in the GeV data have been found to date \citep{Dai2020}. 

\begin{figure}
\centering
 \includegraphics[height=5.5cm, width=8.5cm]{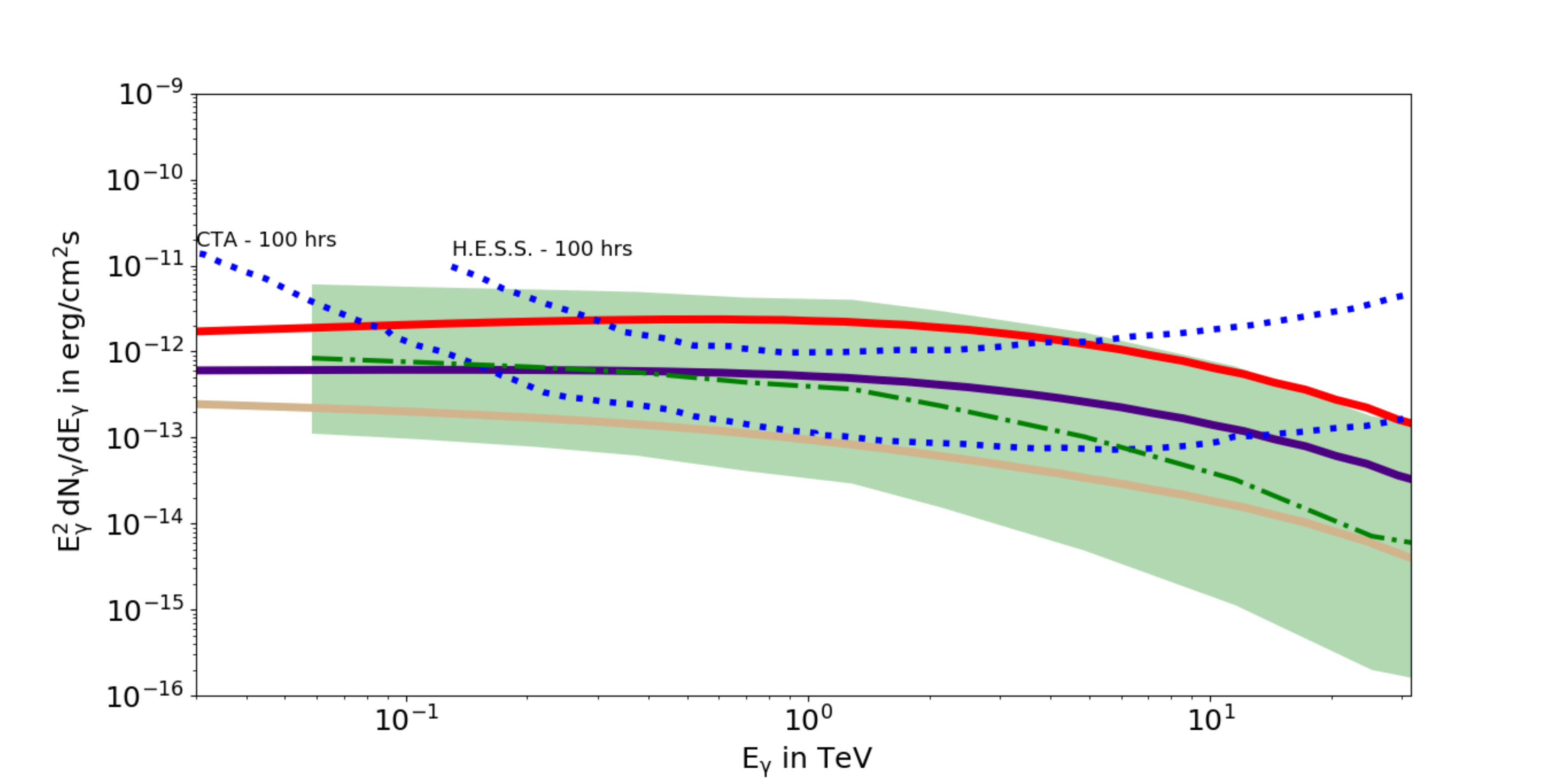}
 \caption{Differential flux predictions (green error band) for $\omega$~Cen, the solid lines indicating example predictions. The blue dashed lines represent the H.E.S.S.\ and CTA sensivity curves for 100 hours of observations \citep{Funk2013}. For the example spectra, we chose $E_{\rm min} = 0.001\,{\rm TeV}$ and $E_{\rm max} = 100\,{\rm TeV}$ for all cases. We then assume Bohm diffusion, $B = 5\mu G, \Gamma = 1.9$, and $Q_{0} = 8.17\times10^{33}\,{\rm erg^{-1}\,s^{-1}}$ (red line); $\kappa_{0} = 7\times10^{-5}\,{\rm kpc^{2}/Myr}, B = 3\mu G, \Gamma = 2.1$, and $Q_{0} = 2.40\times10^{33}\,{\rm erg^{-1}\,s^{-1}}$ (violet line) and $k_{0} = 1.0\times10^{-3}\,{\rm kpc^{2}/Myr}, B = 1\mu G, \Gamma = 2.3$, and $Q_{0} = 7.57\times10^{32}\,{\rm erg^{-1}\,s^{-1}}$ (golden line).}
 \label{fig:wCen}
\end{figure}

We apply the same method as in Section~\ref{MAGIC_method} on $\omega$~Cen as an example of a Southern source. In Figure~\ref{fig:wCen} we compare the predicted differential fluxes with the H.E.S.S.\ and CTA sensitivity curves for 100 hours of observations. We also indicate typical predicted $\gamma$-ray spectra for a few choices of parameter combinations. We note that our current IC predictions are higher than predicted by \citet{Ndiyavala2018}. This is because \citet{Ndiyavala2018} assumed relatively low numbers of stars within this cluster, while we now use $N_*=10^7$ \citep{Miller2019}, creating a much higher ambient target photon density.

\section{Discussion and Conclusions}
A number of models attempt to explain the multi-wavelength spectra detected from several Galactic GCs. In addition to the broad leptonic and hadronic scenarios described in Section~\ref{Introduction}, there are other processes or refinements that we did not consider in this paper, but that can significantly influence the level of the predicted TeV emission from GCs. For example, \citet{Bednarek2014} performed numerical calculations of the asymmetric TeV $\gamma$-ray emission morphology produced by the interaction of stellar winds leaving the GC and the Galactic medium, which should create a bow shock nebula around GC. Furthermore, \citet{Bednarek2016} considered the role of advection of leptons within the GC by the mixture of winds from the embedded populations of
MSPs and other stellar members, as well as the effect of a non-central location of a dominating, energetic MSP, on the predicted $\gamma$-ray flux.
In order to constrain or discard some of these models, careful testing of them is needed. 

This paper focused on assessing uncertainties in the predicted VHE $\gamma$-ray flux of GCs within the context of the leptonic GC model of \citet{Kopp2013}, with the main aim to give theoretical guidance to CTA's observational strategy. We demonstrated that uncertainty in model parameters leads to a large spread in the predicted flux.
We modelled the IC $\gamma$-ray flux expected from 15 GCs that have been observed, but not detected, by H.E.S.S.\ in the VHE band.
We confirmed that a finer grid in parameter space leads to a smoother flux distribution, as expected. Furthermore, a larger number of trials led to convergence of the flux distribution, as $N_{\rm t}\rightarrow N_{\rm comb}$. Also, the eventual predicted range of fluxes depends on the number of free model parameters, as well as their respective ranges. 

Interestingly, we could demonstrate the power of using a stacking approach: the relative errors on the predicted live-time-weighted flux is nearly one order of magnitude lower than that for the individual GCs. This well-known effect is also apparent in the observational upper limits, where the weighted-flux upper limit is lower and therefore more constraining than the single-GC ones (Figure~\ref{fig:UL}). We also note that the median flux and geometric mean flux were very close. The different methods of estimating the error on the mean flux yielded slightly different, yet broadly consistent results, but one can appreciate the large spread (uncertainty) in predicted fluxes based on these error estimates.

We found that while none of our predicted individual cluster median fluxes plus uncertainties violated the respective upper limits (Figure~\ref{fig:UL}), our total integral weighted flux violated the H.E.S.S.\ upper limit for our first set of parameter combinations (Table~\ref{table:1}). However, one has to be careful to summarily discard models if calculations for best-guess parameters exceed observational limits, noting the considerable uncertainty in predictions that may stem from uncertain parameter values. 
%This also illustrates the point that one should not be overly pessimistic when choosing model parameters, as this may render the predictions nearly unusable given the unreasonably high uncertainty in the calculated values. 
%There are parameter combinations that provide fluxes below this stringent limit. 
Our second parameter combination  satisfied the stacked upper limits (Table~\ref{table:2}). Thus we could use this stringent upper limit to constrain the source properties of the MSPs embedded within the GC.
%, to see whether the predicted flux exceeded the CTA sensitivity and also to see if we could satisfy the H.E.S.S.\ upper limits on these sources since there exist upper limits for these clusters. 
%We however expect the error band to increase if we use more free parameters, such as freeing the diffusion coefficient. 
For M15, we found that the differential upper limits were more constraining than the H.E.S.S.\ integral flux upper limits. Yet, we could satisfy these upper limits for typical parameters. As an example of a Southern-hemisphere source, we calculated the TeV flux for $\omega$~Cen, indicating that this source may be a possible candidate to be observed by H.E.S.S.\ or CTA. 

In conclusion, we note that increasing measurement accuracy on model parameters will improve predictions of GC fluxes, and this will provide  better guidance to CTA's observations. Future refined models will have to undergo continued scrutiny, also taking into account the effect of parameter uncertainty on their predictions, as they are confronted with new data.

\section{Acknowledgments}
The Virtual Institute for Scientific Computing and Artificial Intelligence (VI-SCAI) is gratefully acknowledged for operating the High Performance Computing (HPC) cluster at the University of Namibia (UNAM). VI-SCAI is partly funded through a UNAM internal research grant. We also acknowledge discussions with James Allison and Leonard Santana. This study was financially supported by the African German Network of Excellence in Science (AGNES), through the ``Programme Advocating Women in Science, Technology, Engineering and Mathematics''. This work is based on the research supported wholly / in part by the National Research Foundation of South Africa (NRF; Grant Numbers 87613, 90822, 92860, 93278, and 99072). The Grantholder acknowledges that opinions, findings and conclusions or recommendations expressed in any publication generated by the NRF supported research is that of the author(s), and that the NRF accepts no liability whatsoever in this regard.

\section{Data availability}
Data available on request - The data underlying this article will be shared on reasonable request to the corresponding author.
\bibliographystyle{mnras}
\bibliography{ref}
%%%%%%%%%%%%%%%%%%%%%%%%%%%%%%%%%%%%%%%%%%%%%%%%%%
\bsp	% typesetting comment
\label{lastpage}
\end{document}